\documentstyle[11pt,aaspp]{article}
\begin{document}

\title{Lithium-Beryllium-Boron and Oxygen in the early Galaxy}

\author{Elisabeth Vangioni-Flam}

\affil{Institut d'Astrophysique de Paris, CNRS, 
98 bis Boulevard Arago, 75014, Paris, France}
\author{ and Michel Cass\'e}
\affil{Service d'Astrophysique, DAPNIA, CEA
Orme des Merisiers, 91191 Gif sur Yvette, CEDEX, France \\ and 
Institut d'Astrophysique de Paris, CNRS, 
98 bis Boulevard Arago, 75014, Paris, France}

\begin{abstract}
Oxygen is a much better evolutionary index than iron to follow the history
 of Lithium-Beryllium-Boron (LiBeB) since it is the main producer of these
 light elements at least in the early Galaxy. The O-Fe relation is crucial
 to the determination of the exact physical process responsible for the LiBeB
 production. Calculated nucleosynthetic yields
 of massive stars, estimates of the energy cost of Be production, and above all recent observations
 reported in this meeting seem to favor a mechanism in which fast nuclei
 enriched into He, C and O arising from supernovae are accelerated in superbubbles and fragment
 on H and He in the interstellar medium. 
\end{abstract}

\section{Introduction}

Lithium-Beryllium-Boron take a special status in the general framework of nucleosynthesis.
These nuclei, are indeed of exceptional fragility and they are 
destroyed in stars above about 1 million degrees.
The main formation process available is spallation, i.e. 
fragmentation of medium light isotopes (CNO) leading to lighter 
species as ${^6}Li$, ${^7}Li$, ${^9}Be$, $^{10}B$ and $^{11}B$.
The physical parameters  of the spallation mechanism are fourfold:
i) the production cross sections as a function of energy  are fairly well measured in the 
laboratory
ii) the source composition of the energetic component
ii) the associated (injection) energy spectrum
iv) the target composition.
In the following, we describe the two reverse spallative processes able to produce LiBeB
  and we confront them to observational constraints. We show how the 
 relation between oxygen and iron, in the halo
 phase is determining to discriminate between the two processes.

\section{Spallation processes and astrophysical sites}

The pioneering article of 
Meneguzzi, Audouze and Reeves (1971) offered the first 
quantitative explanation of the local abundances of 
LiBeB, at a time when only cumulated abundances in the solar system 
were available.
The standard Galactic Cosmic Rays (GCR), essentially composed of fast protons and alphas collide with 
CNO nuclei sitting in the interstellar medium to yield measured LiBeB abundances. 
But the observed isotopic ratios of Li and B were not reproduced. A stellar 
source of $^{7}Li$ was made necessary. For $^{11}B$  a complementary had hoc
spallative source of low energy was invoqued.

Introducing the time parameter, i.e. taking into account the fact that the amount
 of CNO varies in the ISM  as well as the flux of cosmic rays (protons and alphas), presumably
  like the rate of supernovae, itself responsible for the increment of metallicity, leads to
 the following evolution:
 the abundance of a given light element
 increases like the square of the CNO abundance (or as a good approximation to O).
  Thus according to the 
classical tradition in galactic evolution, the production of LiBeB by 
the GCR is called "secondary".

In the nineties, measurements of Be/H and B/H from KECK and HST, together 
with [Fe/H] in low metallicity halo stars came to set 
strong constraints on the origin and evolution of LiBeB isotopes.
 The evolution of BeB was suddenly uncovered  over about 10 Gyr, taking [Fe/H] 
as an evolutionary index.
The linearity between Be, B and iron came as a surprise since a quadratic 
relation was expected from the standard GCR mechanism. It was a strong
 indication that the standard GCRs are not the main producers of LiBeB in
 the early 
Galaxy. A new mechanism of primary nature (production rate independent of the
 interstellar metallicity, i.e. BeB abundances prop to O) was required to
 reproduce these observations: it has been proposed that low energy CO nuclei ( a few tens MeV/n)
 produced and accelerated by massive stars (WR and SN II in superbubbles, i.e. cavities in
 the interstellar medium excavated by the winds and explosion of massive stars) fragment
  on H and He
 at rest in the ISM. This low energy component (LEC) has the advantage of
 coproducing Be and B in good agreement with the ratio observed in Pop II
 stars (Vangioni-Flam et al 2000 and Ramaty et al 2000).

The term "metallicity", up to now has been ambiguously
 defined. In fact, all the above argumentation
assumes that CNO/Fe is constant at [Fe/H] less than -1.
Since oxygen is the main progenitor of BeB, the
 apparent linear relation between BeB and Fe could be misleading if O was
 not strictly proportional to Fe. Thus the pure primary origin
 of BeB in the early Galaxy could be questionned.

If O/Fe is constant, the GCR process is 
unsignificant due to the paucity of the ISM in CNO in the halo phase. The LEC process is 
obviously predominent, since it is free from the ISM metallicity, 
relying on freshly synthesized He, C, and O. Progressively, 
following the general enrichment of the ISM, GCR gain importance.
Neutrino spallation plays its marginal role, increasing the abundance 
of $^{11}B$.

Recent observations of Israelian et al (1998) and Boesgaard et al (1999) (IB)   
showing a neat slope in the O-Fe plot have seeded a trouble. If 
they are verified, the whole interpretation has to be modified, giving a larger role
 to the secondary process in the halo phase (Fields and Olive 1999). Nevertheless,
 the primary component is also required
 in the very early Galaxy. However 
these observations are considered as controversial, and the whole 
session is centered on this point.
What is the right O-Fe correlation? This point is crucial to translate 
the Be, B-Fe relations into B, Be-O ones.
In the IB observations, [O/Fe]= -0.35 [Fe/H], and consequently log (Be/H)
 is  proportional to 1.55 [O/H]. Fields and 
Olive (1999) use this relation to rehabilitate the classical standard 
 GCR as the progenitor of LiBeB. But, beyond the observational questioning, this scenario
 meets with theoretical difficulties. The 
energy cost to produce a single Be nucleus is unfavorable but not prohibitive since this
 cost is plagued by large uncertainties (Fields et al 2000).
 The main difficulty is that the stellar supernova yields integrated in 
a galactic evolutionary model cannot fit the new O/Fe data (F. Matteucci this meeting).
Moreover, the observational [$\alpha$,/Fe] vs [Fe/H] where $\alpha$ = Mg, Si, Ca, Ti show a plateau
 from [Fe/H] = -4 to -1. It would be surprising that oxygen does not follow the Mg, Si and Ca trends 
 since all these nuclei are produced by the same massive stars.
If the trend expressed in this meeting, namely that [O/Fe] is approximatively 
constant or slightly varying
 in the halo phase is confirmed, then the theoretical situation is clarified and the primary component
 made necessary in the early Galaxy, until at least [Fe/H]= -1.

\section{Conclusion}

 The synthesis of LiBeB in the halo proceeds through nuclear spallation, essentially by
 the break up of oxygen and the $\alpha + \alpha$ reaction.
 The observed relation between BeB and Fe has to be translated through
 the O - Fe relation into a BeB - O which is representative of the relevant physical production
 process. And then the observational O - Fe relation is determining. 
(except if Be, B and O are measured simultaneoulsy in 
stars  (A.M. Boesgaard and K. Cunha, this 
meeting). Anyway the primary component is required in the halo phase, then afterwards the secondary
 process takes over; the question is therefore
 at which metallicity the transition occurs, the answer depends of the behaviour of [O/Fe].
  The ideal test of the scenarios would be to get the evaluation of
 O, Fe, Mg Be, B, Li abundances in the same stars. Impressive progress are waited
 from the VLT. 
Gamma ray line astronomy, through the european INTEGRAL 
satellite, to be lauched in 2002, will help to constrain the energy 
spectrum and intensity of LEC.

\end{document}